\documentclass[aps,prd,preprint,groupedaddress]{revtex4-1}
\usepackage{latexsym,amsfonts}
\usepackage{hyperref}
\begin{document}     
%%%%%%%%%%%%%%%%%%%%%%%%%%%%%%%%%%%%

\title{Finite temperature quantum field theory in the heat kernel method
\footnote{\href{http://dx.doi.org/10.1134/S1061920815010033}{Russian Journal of Mathematical Physics, {\bf 22} (1), 9-19 (2015)},\\
\href{http://hdl.handle.net/11858/00-001M-0000-0025-0ADC-A}{hdl.handle.net/11858/00-001M-0000-0025-0ADC-A}}}

\author{Yuri V. Gusev}

\affiliation{Lebedev Research Center in Physics, Leninskii Prospekt 53, Moscow 119991, Russia}
\affiliation{Max Planck Institute for Gravitational Physics (Albert Einstein Institute), Am M\"uhlenberg 1, D-14476 Golm, Germany}
\affiliation{Institut des Hautes \'Etudes Scientifiques, Le Bois-Marie, 35 route de Chartres, 91440 Bures-sur-Yvette, France}

\date{September 26, 2014}

\begin{abstract}
The  trace of the heat kernel  in a {(D+1)-dimensional} Euclidean spacetime (integer D $\ge$ 2) is used to derive the free energy in finite temperature field theory. The spacetime presents a {D-dimensional} compact space (domain) with a (D-1)-dimensional boundary, and a closed dimension, whose volume is proportional the Planck's inverse temperature. The thermal sum appears due to topology of the closed Euclidean time. The obtained free energy in {(3+1)} and {(2+1)} dimensions contain two contributions defined by the  volume of a domain and by the volume of the domain's boundary. This functional is finite and valid for arbitrary values of the Planck's inverse temperature. The absolute zero of thermodynamic temperature is forbidden topologically, and no universal low temperature  asymptotics of the free energy can exist.
\end{abstract}
\maketitle

%%%%%%%%%%%%%%%%%%%%%%%%%%%%%%%%%%%%		
%\section{Motivation}      \label{intro}                                        
%%%%%%%%%%%%%%%%%%%%%%%%%%%%%%%%%%%%

Quantum field theory (QFT) at finite temperature  is a subject developed for long time \cite{Vasiliev-book1998,Kapusta-book2006,Umezawa-book1982,Hurt-book1983}, nevertheless it needs some clarifications, as emphasized by recent experimental observations in the lower dimensional condensed matter physics and in the heavy-ion collision experiments of collider physics.
Let us begin with a trivial statement that different physical theories may not overlap, and one theory is not necessarily a limiting case of another. This is certainly the case with quantum statistics and finite temperature quantum field theory, and these two should not be mixed up. Therefore, we have to be careful with terminology since same names are sometimes used for entities that belong to different theories, even though these entities may not be equivalent, e.g. free energy. There are also some technical problems with finite temperature QFT. One problem is that the method imported from the study of scattering problems in accelerator physics, namely the Feynman diagram expansion based on the Green's functions  \cite{Abrikosov-book1965}, is not the best for finite temperature QFT. The Feynman path integral and the thermal Green's functions are formulated in phase space, but the phase space formalism is derived and valid for physics of massive particles.  Known particles and quasiparticles in condensed matter physics \cite{Fetter-book2003}, where models start with Hamiltonians, used to be massive until the situation has changed with the recent discoveries of massless quasiparticles \cite{Novoselov-Nature2005}. However, massless particles cannot be localized \cite{Schwinger-book1989}, and there is no threshold on the production of massless particles \cite{GAV-lectures2007}, so the meaning of traditional physical notions (mean free path, particle density, etc.) gets lost. In addition, the Feynman integrals in massless QFTs suffer from the infrared (IR) divergences caused by calculation methods. These IR divergences overlap with the standard ultraviolet (UV) divergences of relativistic quantum theories, and as a consequence finite temperature QFT becomes almost intractable \cite{Kapusta-book2006}. Introducing mass-like physical parameters (particles' masses and chemical potentials), whose values are calibrated by experimental data, may help getting rid of the divergences and eventually fitting observations, but the predictive power of such theories may be limited.

To resolve these problems we suggest to use QFT techniques, which originate from Julian Schwinger's ideas \cite{Schwinger-PR1951,Schwinger-talk1973,Schwinger-book1989} of implementing the Lagrangian formalism in spacetime.  The method of effective action, which is based on the Schwinger-DeWitt (geometrical) QFT formalism \cite{DeWitt-book1965,DeWitt-book2003,BarVilk-PRep1985}, is used below for finite temperature field theory. The main functional of finite temperature QFT is {\em free energy}, which is different in its use and meaning from free energy in statistical physics.  The effective action or free energy is a phenomenological functional derived from the heat kernel. In the Schwinger-DeWitt formalism the interactions (gauge fields, gravity) are built-in via geometry. The notion of a mean field is known in condensed matter physics, and its QFT meaning is the field's expectation values \cite{GAV-lectures2007}. 

%%%%%%%%%%%%%%%%%%%%%%%%%%%%%
\section{General  principles}
%%%%%%%%%%%%%%%%%%%%%%%%%%%%%

%%%%%%%%%%%%%%%%%%%%%%%%%%%%%%%%%%%%%%%%%%%%%
\subsection{Temperature}  \label{temperature}
%%%%%%%%%%%%%%%%%%%%%%%%%%%%%%%%%%%%%%%%%%%%%

The key concept of any theory of thermal phenomena is {\em temperature} \cite{Quinn-book1990,Sommerfeld-bookV5}. Temperature is an intrinsic parameter of phenomenological theories from statistical thermodynamics \cite{Kubo-book1968} to the Ginzburg-Landau model of superconductivity \cite{Kittel-book1996}. One usually understands the temperature as a measure of the total energy density of electromagnetic interactions among constituents (particles) of a condensed matter system \cite{Landau-V5}. This energy can be stored in or transferred out of a condensed matter system \cite{Fourier-book1878}. Since there are several kinds of constituents in condensed matter, e.g., elastic and electronic heat contributions, the unique temperature for the whole system in principle is not guaranteed. Different kinds of energy should be convertible to each other to have the unique temperature in a system for otherwise more than one temperature would exist, which is not this case considered here.   

The temperature of gases is due kinetic energy of  massive molecules, thus it is physically different from temperature of condensed matter where particles do not freely propagate, but rather form a continuous medium. Temperature is measured by a thermometer, however, a thermometer does not really measure the energy density, but rather the power flux from condensed matter medium through their common boundary. We assign a single constant value to the temperature of a whole condensed matter system. This may be considered similar, even though not equivalent, to thermal equilibrium in statistical thermodynamics. Later we will have to relax this restriction. Temperature is strictly positive, $T>0$, i.e., zero absolute temperature cannot occur in matter. Mathematical justification of this fact is given below geometrically, while physically it corresponds to the empirical law of thermodynamics that one cannot attain the absolute zero of temperature. In textbooks, this statement follows from the third law of thermodynamics (the Nernst-Planck theorem) \cite{Kubo-book1968} and is based of the notion of entropy. We hope that explicitly including the physical system's topology into  theory, as done in the heat kernel method, one could mitigate some problems with the use of entropy. 

Temperature is one of input variables of the theory, and let us introduce it according to the principle previously found in QFT. Since any condensed matter system exists in some spatial $D$-dimensional domain and its behaviour in time is not studied, we can set up a field theory in the Euclidean $(D+1)$-dimensional spacetime and then declare one of the dimensions closed, $\mathbb{S}^1$. This procedure allows us to keep the correct number of spacetime dimensions, while removes time from the spacetime variables. We now identify  the length (volume) of the closed Euclidean dimension with the inverse temperature expressed in Planck's natural units \cite{Planck-book1914}. Thus, the  {\em Planck's inverse temperature}, with the fundamental constants explicitly present, is, 
\begin{equation}
\beta = \frac{\hbar v}{m T k_B}, \label{beta}
\end{equation}
where $k_B$ is the Boltzmann's constant, $\hbar$ is the Planck's constant. This is now the true variable of the theory, there is nor temperature, neither time, only $\beta$. The characteristic velocity parameter $v$ enters (\ref{beta}), for the electronic component of heat it is the speed of light $c$, while for elastic waves - the velocity of sound.  The value of the calibrating number $m$ in (\ref{beta}) is defined from experiments.  The question of what is fundamental here, the temperature $T$ or the closed time length parameter $\beta$, resembles a situation in the gravity theory, where matter defines geometry (metric) through the energy-momentum tensor. Here too, matter defines geometry ($\beta$) through energy ($T$). This means that for different energy contributions (elastic, electronic) there are different geometrical parameters $\beta_{i}$. The question will be studied in more detail in the theory of thermal and electronic properties of matter.

The definition (\ref{beta}) is common in the literature, with the standard choices of $m=1$ and $v=c$. This definition of the Planck's inverse temperature is different from the common definition $\beta' =1/(T {k_{\mathsf{B}}})$ used in statistical physics \cite{Kubo-book1978}, which scales as inverse energy $J^{-1}$. The reason for this choice of $\beta'$  was to make energy dimensionless in the distribution functions because discrete energy states are explicitly used in statistical mechanics. In a field theory, the only dimensionful quantity is {\em length} that calls for the expression (\ref{beta}). Parameter $\beta$ gives the characteristic length at a given temperature, the thermal wavelength, which at room temperature is about $3.8\ \mu m$.

%%%%%%%%%%%%%%%%%%%%%%%%%%%%%%%%%%%%%%
\subsection{Laplacian}  \label{Laplacian}
%%%%%%%%%%%%%%%%%%%%%%%%%%%%%%%%%%%%%

In quantum field theory, one usually starts with the Lagrangian which defines the action of a theory (the Lagrangian approach began with P.A.M. Dirac' work \cite{Dirac-PZS1933}). By making two variation derivatives, the Hessian can be derived to obtain the Laplace operator from the theory's action. The Laplacian is the fundamental object of any field theory, and instead of appealing to the action, one could {\em define} a field theory by its Laplacian. Most physically relevant theories can be specified by the Laplace type operator \cite{BarVilk-PRep1985},
\begin{equation}
     \hat{F}(\nabla)= \Box \hat{1} +  \hat{P}(x),
     \label{operator}  
\end{equation}
which satisfies the equation on the field $\varphi$ of arbitrary spin-tensor structure,
\begin{equation}
     \hat{F}(\nabla) \varphi = 0.
\end{equation}
The Laplacian (Laplace-Beltrami operator) is constructed of the covariant derivatives, 
\begin{equation}
\Box= g^{\mu\nu} \nabla_{\mu} \nabla_{\nu}, \label{laplacian}
\end{equation}
that contain the metric and gauge field connections characterized by the commutator curvature, 
\begin{equation}
 (\nabla_{\mu} \nabla_{\nu}-
\nabla_{\nu} \nabla_{\mu} ) \varphi =
\hat{\cal R}_{\mu\nu} \varphi.  \label{commutator}
\end{equation}   
In the Abelian gauge field setting, it is proportional to the Maxwell's electromagnetic tensor, which is not featured below due to an approximation used. The overhat symbol indicates the matrix structures, like in the potential term $\hat{P}={P^A}_B$ that can be an arbitrary local function of the background fields. The scalar Ricci tensor $R$ is usually explicitly present in (\ref{operator}) to work with gravity, we temporarily leave it out of consideration.  Generically, any curvature tensor, potential or field strength is denoted here as $\Re$. Theories that do not fit the class (\ref{operator}), e.g., whose Lagrangians generate the higher order and non-minimal operators, can still be reduced to the given form by algorithms presented in \cite{BarVilk-PRep1985}. The Dirac operator should also be used as the Dirac Laplacian (\ref{laplacian}), as widely studied in the mathematics literature. It was introduced first by Vladimir Fock \cite{Fock-ZfP1926}, and on the Riemannian manifolds by Bryce DeWitt  \cite{DeWitt-book1965}, nevertheless it is known as the Lichnerowicz operator.

This is all we need to know about the type of field theories. This is the setup developed in the effective action method. It is very general and applicable to different areas of physics including ordinary thermal phenomena.

%%%%%%%%%%%%%%%%%%%%%%%%%%%%%%%%%%%%%%%%%%%%%%%%%%
\subsection{Topology}
%%%%%%%%%%%%%%%%%%%%%%%%%%%%%%%%%%%%%%%%%%%%%%%%%%

The key to building a condensed matter theory is topology, let us study a condensed matter system as a field theory on a compact domain $\mathcal{M}^{D}$ of the space $\mathbb{R}^{D}$, with a boundary ${\mathcal{B}}^{D-1}$. The effects of boundary and edges of a spatial domain that the condensed matter is confined to should be accounted for, because the boundary really defines condensed matter. From experiments we know that effects of the size and the boundary of a condensed matter system on its physics can be large, sometimes they can be the leading contributions. The geometrical formalism based on the heat kernel naturally incorporates a boundary as shown in the next section. Topology of the closed Euclidean time $\mathbb{S}^1$ is already mentioned in the temperature definition. 

In the geometrical thermodynamics, where the only scale is a length, the field theory's scale is restricted at the upper length limit and may be restricted at the lower length limit. The upper limit is due to a finite size of the system, it is naturally taken into account by a method of spectral geometry we use: the heat kernel trace. In condensed matter, the lower limit for the elastic energy  contribution is introduced by the size of a lattice cell; this is is the limit of validity of the continuous medium model. 

%%%%%%%%%%%%%%%%%%%%%%%%%%%%%%%%%%%%%%%%%
\section{Heat kernel}  \label{HK}
%%%%%%%%%%%%%%%%%%%%%%%%%%%%%%%%%%%%%%%%

The solutions of the heat equation,
\begin{equation}
	\Big(\frac{\partial}{\partial s} - \Box^x \Big) K (s| x,x') = 0,\ 
	K (s| x, x')|_{s\rightarrow 0} = \delta(x,x'),         \label{heateq}
\end{equation}  
can be applied to theories of fields of different nature \cite{Vilkov-Strasb}.  The parameter $s$ is called the {\em proper time} and it has the dimensionality of the square meter.  Its use is due to Fock \cite{Fock-pt1937}, but it became popular after works of Schwinger \cite{Schwinger-PR1951}. The heat equation's name originated from the heat theory of J.B.J. Fourier \cite{Fourier-book1878}, however, no temperature enters Eq.~(\ref{heateq}), and $s$ is not the time. The Fourier ('thermal') heat equation is the diffusion equation with the partial derivative over the physical time \cite{Vladimirov-book1971}. Contrary, the proper time is a parameter in the  {\em evolution} equation (\ref{heateq}) that is set up for the kernel $K(s|x,x')$, which is a two-point functional.

In the context of field theory, the heat equation is used to obtain the kernel of a field operator (\ref{operator}).  The fundamental solution for the heat kernel \cite{BarVilk-PRep1985,DeWitt-book2003},
\begin{equation}
\hat{K}(s| x, x') = \frac{1}{(4 \pi s)^{D/2}} \mathcal{D}^{1/2}(x,x') 
\exp\Big(  -\frac{\sigma(x,x')}{2s} \Big) \hat{a}_0(x, x'), \label{K0}
\end{equation}
is expressed through the world function $\sigma(x,x')$, which was introduced by Harold Ruse \cite{Ruse-PLMS1931} and made a working tool of general relativity by John Synge \cite{Synge-book1960}. The two-point world function is half a square of the geodesic distance between spacetime points $x$ and $x'$ \cite{Synge-book1960,DeWitt-book2003}.  $\hat{a}_0(x, x')$ in (\ref{K0}) is the parallel transport operator \cite{BarVilk-PRep1985}, needed to transport the indexes of internal degrees of freedom. Below it would only introduce  the matrix trace $\mathrm{tr}\hat{1}$ in the final result. Finally, $\mathcal{D}(x,x') $  is the van Vleck-Morette determinant \cite{DeWitt-book2003} whose coincidence limit gives the metric's determinant $g(x)$. Only the prefactor and the world function's exponent depend on the proper time and will be essentially used in our derivations. Starting from the fundamental solution, several computational techniques for the heat kernel can be worked out \cite{Avramidi-RMP1999}, e.g., the Schwinger-Dewitt (short proper time) expansion \cite{DeWitt-book1965,BarVilk-PRep1985}, the Barvinsky-Vilkovisky (covariant perturbation) theory \cite{CPT1,CPT2}, etc. Our problem is insensitive to a computational method in the approximation below.

It was shown \cite{DeWitt-book1965,BarVilk-PRep1985} that the effective action can be computed through the heat kernel. In fact, to derive QFT free energy, which is a finite temperature equivalent of the effective action, one needs only the functional trace of the heat kernel,
\begin{equation}
{\textrm{Tr}} K (s)  =
\int {\mathrm d}^{D+1}  x \,  {\textrm{tr}}\,  \hat{K} (s|x,x). \label{TrK}
\end{equation}
It has the matrix trace over internal indexes $tr$, assumes the coincidence of spacetime points and integration over the whole spacetime. The density of the integration measure $g^{1/2}(x)$ is included in $K(s|x,x)$. Obviously, ${\textrm{Tr}} K (s)$ is a dimensionless functional, in contrast to the heat kernel (\ref{K0}). The zeroth order of the heat kernel trace is just the functional trace of the fundamental solution (\ref{K0}) \cite{BarVilk-PRep1985,CPT2}.   

Computational techniques for the heat kernel typically assume open, asymptotically flat spacetimes, i.e., $\mathbb{R}^D$ manifolds. But here we are interested in compact spaces with boundaries, the simplest example could be the three-dimensional interior with a two-dimensional boundary. The heat kernel trace $\mathrm{Tr}K(s)$ in the compact manifold $\mathcal{M}^D$ with the boundary ${\mathcal{B}}^{D-1}$ can be calculated in the covariant perturbation theory \cite{CPT2}.  It is easy to write down fundamental solutions for the heat kernel trace on $\mathcal{M}$ and on $\mathcal{B}$, which are defined by volume of the domain and by volume (area) of the boundary.  Then we have the following truncated series for the kernel of the heat equation in a compact manifold with boundary,
\begin{equation} 
{\mathrm{Tr}} K(s) = 
	\frac1{(4\pi s)^{D/2}}\ \mathcal{V}\
 	{\mathrm{tr}}\,   \hat{1}
 	+ \frac1{(4\pi s)^{(D-1)/2}}\
 	\mathcal{S}  \,
 	{\mathrm{tr}}\hat{1} 
 	 + \mathrm{O}[\Re].  \label{TrKlocal} 
\end{equation}
This expression is valid at arbitrary proper time values, i.e., it is {\em not} a short proper time expansion. Here the volume of the $D$-dimensional domain is denoted $\mathcal{V}$ and the area (in mathematics literature both of these two terms are referred to as volume, we use the term 'area' of the convenience of physicists) of its $(D-1)$-dimensional smooth boundary is denoted $\mathcal{S}$. The formal definitions are,
\begin{equation} 
	\mathcal{V} = \int_{\mathcal{M}} \mathrm{d}^{\mathrm{D}} x g^{1/2} (x),
\end{equation}
\begin{equation} 
 	\mathcal{S}  = \int_{{\mathcal{B}}} \mathrm{d}^{\mathrm{D-1}} x \bar{g}^{1/2} (x),
\end{equation}
where $\bar{g}$ is the metric determinant of the boundary manifold. This is a covariant result even though the curvatures and field strengths do no appear in (\ref{TrKlocal}) explicitly. However, they are present in the heat trace remainder $\mathrm{O}[\Re]$.

It used to be popular \cite{DeWitt-book2003,BarVilk-PRep1985} to study the short proper time  expansion around the exact (\ref{TrKlocal}). This expansion is called the Schwinger-DeWitt expansion in physics literature  and it is really a series in powers of the dimensionless combination $(\Re s \ldots \Re s)$, where $\Re$ denotes any curvature tensor, because the $s \rightarrow 0$ expansion is not allowed. The heat kernel  is nonlocal starting from the first order in the field strength \cite{CPT2,Gusev-NPB2009}. This expression has to  be known at arbitrary proper times, because the short proper time expansion  is not acceptable in compact domains. It is the late time (also called intermediate in other areas of physics) asymptotics \cite{CPT2,BGVZ-JMP1994asymp} that one seeks.

%%%%%%%%%%%%%%%%%%%%%%%%%%%%%%%%%%%%%%%%%%%%%%%%%%%%%%%%%%%%%
\section{Free energy in finite temperature field theory}  \label{FTQFT}
%%%%%%%%%%%%%%%%%%%%%%%%%%%%%%%%%%%%%%%%%%%%%%%%%

The concept of effective action was first introduced  in quantum electrodynamics (QED)  by Julian Schwinger \cite{Schwinger-PR1951,Schwinger-book1989}. Schwinger also pioneered the Euclidean spacetime formalism \cite{Schwinger-PNAS1958}, which is the natural setting for QFT \cite{CPT1}. The one-loop  effective action ('one-loop' name being historical, as there is no Feynman diagrams expansion here) embodies all relevant information about a field theory \cite{DeWitt-book2003}.  The covariant effective action via the heat kernel in spacetime was developed by Bryce DeWitt \cite{DeWitt-book1965,DeWitt-book2003} in order to extend the method to gravity and gauge field theories. In the Schwinger-DeWitt formalism, quantum functionals are expressed in terms of mean (expectation value) fields. This covariant effective action theory had been greatly advanced by Grigori Vilkovisky as outlined in \cite{GAV-Gospel,GAV-lectures2007}.  More field theory definitions and explanations can be found in \cite{DeWitt-book2003,BarVilk-PRep1985}. 

In the Schwinger-DeWitt formalism, one starts with the Euclidean spacetime with $D+1$ dimension and specifies its topology. We assume that the Euclidean time dimension is the closed manifold ${\mathbb{S}}^1$. The length of an orbit of ${\mathbb{S}}^1$ is denoted by $\beta$. We make its product with the $D$-dimensional space, $\mathbb{R}^{D} \times {\mathbb{S}}^1$, and study geodesics in this spacetime. We cannot use the term 'world line' because it is associated with the geodesic trajectory of a particle.  In this geometrical picture without particles, there is no direction of the geodesic, we only look for solutions of the differential equation (\ref{heateq}).

To compute $\mathrm{Tr} K^{\beta}(s)$ in the prescribed geometry, the heat kernel could be (under conditions specified below) factorized into the Euclidean time heat kernel,  which is the one-dimensional flat (\ref{K0}), and the spatial part $\mathrm{Tr} K^{(D)}(s)$.  The procedure for computations is to make a geodesic loop and then shrink it to a point, this is the coincidence limit $x=x'$ \cite{DeWitt-book1965,DeWitt-book2003,BarVilk-PRep1985}. Here this cannot be done because topology of $\mathbb{S}^1$ prevents it. There is an incontractible geodesic loop (which topologically translates to a hole), and any number of loops is allowed that brings the sum over $n=1, \ldots , \infty$.  The world function in the one-dimensional coordinate $\tau$ is trivially half a square of the geodesic length,
\begin{equation} 
\sigma(\tau, \tau') = (\tau-\tau')^2 /2.
\end{equation} 
The length of the closed geodesic is proportional to $\beta$, and the factor $n$ counts the windings. Therefore, using Eq.~(\ref{K0}) we can write down the heat kernel trace in the given spacetime as,
\begin{equation} 
	{\textrm{Tr}}{K}^{\beta}(s)=  
	\frac{\beta}{(4 \pi s)^{1/2}}\
	\sum_{n=1}^{\infty} {\mathrm{e}}^{-\frac{\beta^2 n^2}{4s}}  \
	\int {\mathrm d}^{D} x \,   
	{\mathrm{tr}}\, {K}^{(D)} 
	(s|x,x),  \ D \ge 2.  \label{TrKbeta}
\end{equation}

Now we {\em define} the functional of free energy $F^{\beta}$ as the proper time integral of the trace of the heat kernel obtained above,
\begin{equation}
-F^{\beta} \equiv
 \int_0^{\infty}\! \frac{{\mathrm d} s}{s}\,  
	{\mathrm{Tr}}  K^{\beta}(s).         \label{Fbeta}
\end{equation}  
This proper time integral is known in mathematics as the Mellin transform. The functional (\ref{Fbeta}) is dimensionless because both the integrated functional and the proper time integral are dimensionless. The definition for the effective action \cite{BGVZ-NPB1995} is not different from (\ref{Fbeta}), only topology of its spacetime is different. Any available technique can supply a solution for $\mathrm{Tr}\, {K}^{(D)} (s)$, but the heat kernel trace should be known at arbitrary proper time in order to generate $F^{\beta}$ valid at arbitrary $\beta$. 

Computing (\ref{Fbeta}) with the heat kernel trace (\ref{TrKlocal}) in a $(3+1)$-dimensional spacetime is easy. It amounts to substitution of the $D=3$ heat kernel trace (\ref{TrKlocal}) into the finite temperature Eq.~(\ref{TrKbeta}). The subsequent computation of the proper time integral (\ref{Fbeta}) can be done with the new dimensionless variable $y =\beta^2/4 s$. The two terms from $\mathrm{Tr} K^{(D)}(s)$ appear in $F^{\beta}$ as the standard integrals and sums,
\begin{equation}
	\int_{0}^{\infty}{\mathrm{d}} y   \,
	  y^{a-1} \,
	  \sum_{n=1}^{\infty}\, \mathrm{e}^{-y n^2} \,
	=  \zeta (2 a) \,  \Gamma(a),          \label{zetas}
\end{equation}
where $\zeta$ is the Riemann zeta function, and $\Gamma$ is the gamma function.

For three dimensions, the parameter $a$ takes values 2 and 3/2. Then, Eq.~(\ref{zetas}) gives coefficients $\pi^4/90$ and $\zeta(3)\sqrt{\pi}/2$ correspondingly, and we arrive at the following expression,
\begin{equation}
-F^{\beta} = \frac{1}{ \beta^3} \, \frac{\pi^2}{90} \, \mathcal{V}^{(3)}\, \mathrm{tr}\hat{1}  +  \frac{1}{\beta^2}\, \frac{\zeta(3)}{2 \pi}\,  \mathcal{S}^{(3)}\, \mathrm{tr}\hat{1}   
+ \mathrm{O}[\Re], \ D=3. \label{Fenergy}
\end{equation} 
It is easy to get a similar expression for the free energy in (2+1) dimensions,
\begin{equation}
-F^{\beta} = \frac{1}{ \beta^2} \, \frac{\zeta(3)}{2 \pi} \, \mathcal{V}^{(2)}\, \mathrm{tr}\hat{1}  +  \frac{1}{\beta}\, \frac{\pi}{6}\,  \mathcal{S}^{(2)}\, \mathrm{tr}\hat{1} + \mathrm{O}[\Re], \ D=2.  \label{Fenergy2}
\end{equation} 
Here the upper indexes indicate the dimension of the base manifold. In plain words, the boundary's area $ \mathcal{S}^{(2)}$ corresponds to the length of an 'edge' while the volume $\mathcal{V}^{(2)}$ is the 'area'. It is obvious that in two dimensions the free energy (finite temperature effective action) is also (UV and IR) finite. This expression could describe thermal properties of genuine two-dimensional systems, if they were present in Nature, i.e., graphene, even if suspended in vacuum, is still {\em embedded} in a 3D space.
  
The spacetime may have the metric tensor (gravity) and the vector bundle (gauge fields) incorporated via the covariant derivative (\ref{laplacian}). By the proposed algorithm (\ref{Fbeta}), the free energy can be computed only in spacetimes whose gravitational field is specified by ultrastatic metrics, i.e., the global timelike Killing vector exists. The flat spacetime metric is trivial and so ultrastatic. For spacetimes with such geometrical properties, Eq.~(\ref{Fenergy}) is valid at arbitrary temperature, in flat or curved space, for any field theory with the operator (\ref{operator}). This is only a part of a general expression whose other terms depend on gravity curvatures and gauge field strengths explicitly.  The missing terms $\mathrm{O}[\Re]$ are small in the 'high temperature' limit considered below, but they may be not small in general. In the approximation above, the matrix trace $\mathrm{tr}\hat{1}$ may safely be discarded as an unessential numerical factor. This is not the case for other terms not derived here.

This free energy is defined only by the system's global geometrical properties, the space's volume and its boundary's area. These contributions are arranged according to their geometrical properties, not according to their magnitudes. As is seen from (\ref{beta}), the $1/\beta^2$ term may become larger for some physical conditions, or other terms, not displayed, may become dominant. 

Since both sides of (\ref{Fenergy}) are dimensionless, it is meaningless to talk about the high or low temperature limits of free energy in terms of the units of absolute temperature $K$. 
An expansion in $F^{\beta}$ can only be made in a small {\em dimensionless} parameter.
We introduce the {\em effective size} $r$ of a compact space as the ratio of its volume to its boundary's area,
\begin{equation}
 r \equiv \frac{\mathcal{V}}{\mathcal{S}},  \label{effsize}
\end{equation}
e.g., the effective size of a sphere of radius $b$ is $r =b/3$. Then the high temperature limit can be understood as the asymptotic,
\begin{equation}
\beta/r \ll 1. \label{highT}
\end{equation}
This expression places a quantitative restriction on how large a body or a cavity should be compared to its thermodynamic temperature for the high temperature limit to hold.  

Similarly, we could formulate the low temperature limit as, $\beta \gg r$. However, this asymptotic cannot be derived from the displayed expression (\ref{Fenergy}) because it is in $\mathrm{O}[\Re]$.  The low temperature regime,
\begin{equation}
\frac{\hbar c }{k_{\mathsf{B}}} >   r T, \label{lowT}
\end{equation}
holds when the boundary's area energy in $F^{\beta}$ is larger than the volume contribution.

However the {\em true} low temperature asymptotic $\beta \gg   r$ is hidden in the unstudied remainder $\mathrm{O}[\Re]$, which is specific of the physical system's properties (material, surface curvatures, etc.).  The expression (\ref{Fenergy}) is valid only for the 'high temperature' asymptotic (\ref{highT}). The discarded remainder is negligible in this limit, but not in the opposite 'low temperature' one (\ref{lowT}).  A universal low temperature asymptotic for free energy and correspondingly for all its derivative functions does not exist. One would see instead an infinite variety of condensed matter characteristics on the way towards the absolute zero. In particular, we can foresee the ever growing number of quasiparticles being introduced in condensed matter physics (the present stream includes  polaron, exciton, polariton, plasmon, etc). As experimental devices are getting smaller and temperatures lower, physics complexity is increasing. However, this only an {\em asymptotic} behaviour as the limit of the absolute zero temperature, $T \equiv 0$, is forbidden topologically for otherwise topology would have to change from the closed manifold ${\mathbb{S}}^1$ to the open one $\mathbb{R}^1$. 

The expression (\ref{Fenergy}) is found up to a common factor; this constant can only be found from experiments. The name of free energy really masks its true meaning. Free energy is a generating functional for physical observables or effective equations whose solutions can deliver physical observables. It is similar to the classical action that delivers the equations of motion or to the effective action of quantum field theory that generates the effective equations \cite{GAV-AP1998}. The effective equations should be calibrated to the physical observables of our instruments. Physical observables are expressed in the units of the SI system \cite{SI-book}, with help of the fundamental physical constants \cite{CODATA-RMP2012}.  The full structure of free energy expressed via the field strengths, which was discarded in (\ref{TrKlocal}), should be obtained and employed for the study of radiation and electronic phenomena.

%%%%%%%%%%%%%%%%%%%%%%%%%%%%%%%%%%%%%%%%%
\section{Summary}
%%%%%%%%%%%%%%%%%%%%%%%%%%%%%%%%%%%%%%%%%

In order to apply finite temperature quantum field theory to condensed matter physics we had to revisit some of its notions. Even though we only began working on this program, some of its features are apparent.  There is no need to use the thermal fields or the finite temperature Green's functions in  finite temperature QFT. Free energy at arbitrary temperature has no (ultraviolet or infrared) divergences. 

Let us summarize the main points.
\begin{itemize}
\item
The natural variable in thermodynamics is the Planck's inverse temperature. 
\item
The zero absolute temperature limit is topologically forbidden. 
\item
The thermal sum appears from the topology of spacetime. 
\item
Free energy  is a phenomenological functional defined by the spatial domain's geometrical characteristics and the Planck's inverse temperature.
\end{itemize}

%%%%%%%%%%%%%%%%%%%%%%%%%%%%%%%%%%%%%%%%%%%
\section{Discussion}
%%%%%%%%%%%%%%%%%%%%%%%%%%%%%%%%%%%%%%%%%%%

Touching a bit of history, the method we used can be traced to the Kubo-Martin-Schwinger (KMS) condition \cite{Araki-SPTP1978}. This is the condition of periodicity of the finite temperature Green's functions (or field correlators)  in the imaginary, i.e., Euclidean, time. This condition was discovered simultaneously by KMS physicists \cite{Kubo-JPSJ1957-2,Martin-PR1959} and by Efim Fradkin \cite{Fradkin-JETP1959}. KMS condition can be used to express  the 'thermal' Green's function as a sum of the zero temperature Green's functions \cite{Brown-PR1969}. This 'image sum' (thermal sum) is also called the Matsubara sum \cite{Matsubara-PTP1955}. The idea of the 'image sum' technique was later adopted to the heat kernel method  \cite{Dowker-JPA1978,Gusev-PRD1998}. However, it was done too literally  as the zeroth mode is not present in the thermal sum for the heat kernel (\ref{TrKbeta}) as obvious from the definition. This oversight is a legacy of the phase space formalism of the Green's functions, and it caused an apparent UV divergence in the finite temperature QFT, where there is none. 

An advantage of the field theory is that it is a dimensionless theory. Equation (\ref{Fenergy}) is one step away from dimensionless QFT, but it is a necessary step because temperature (\ref{beta}) introduces the scale into a field theory. However, the field theory's functionals remains dimensionless, e.g., $[\beta^3] = [V]$ in (\ref{Fenergy}), because the theory is set up in a finite domain, which gives the natural length scale, (\ref{effsize}). It is obvious that finite temperature field theory can only be defined in compact spaces with boundaries. That is it, a space should have a finite volume for otherwise the spatial integral of the first term is not defined. The second term is defined by the boundary, which is required because physically we aim at describing condensed matter, which is always bounded; mathematically we seek solutions (although implicitly) of the differential equations with certain boundary conditions. 

Since the geometrical formalism uses the spacetime with the Euclidean metric signature, there is no concept of the motion, or the time for that matter. The physical time appears in other formulations of QFT at finite temperature \cite{Kapusta-book2006}, but only at intermediate derivations, while final expressions for physical observables are time independent. The closed Euclidean time method helps us to put a field theory into a non-relativistic form, while keeping the correct number of spacetime dimensions. We accept Roger Penrose's view on physics \cite{Penrose-book1984-1} that the derived algebraic relations are more important than geometrical interpretations that produced them. It means that an existence of non-trivial topology of the spacetime of our theories may be exhibited by the condensed matter or other physical phenomena, but we are only interested in physics that could follow from the derived equations.

No chemical potentials enter finite temperature field theory, because they belong to a different physical theory, the theory of thermodynamic ensembles \cite{Tolman-book1949,Kubo-book1968}. Their presence is not consistent with the principles of quantum field theory. As the name says the chemical potentials come from the thermodynamics of systems with chemical reactions \cite{Prigogine-book1998}. Grand canonical ensemble is used to describe systems in a closed space, in a contact with an external reservoir of particles and energy \cite{Tolman-PR1940,Tolman-book1949,Kubo-book1968}. Chemical potentials denote energy densities of particles supplied to the system from this external reservoir. Therefore, chemical potentials are introduced in order to deal with the variable number of particles. However, quantum field theory by its purpose deals with the creation and annihilation of particles (or the production and backreaction of energy fluxes in the field theory language). Any QFT should have a built-in mechanism for the particle creation {\em and} its backreaction. In the effective action technique, the particle creation mechanism is in the nonlocal effective equations \cite{GAV-lectures2007}. Without such a mechanism but with chemical potentials, a quantum field theory becomes effectively the many body physics.  However, their use is popular \cite{Kapusta-book2006}  because being mass-like the chemical potentials suppress the UV divergences in other formulations of QFT.

The statistical approach to the physics of condensed matter systems has been developed by Victor Maslov  using combinatorics and probability theory \cite{Maslov-RJMP2007}. The 'undistinguishing statistics of objectively distinguishable objects' \cite{Maslov-MN2013} and finite temperature field theory are theories built on different concepts. Maslov answers the question \cite{Maslov-RJMP2007}, {\em ``How to translate measure, density, and dimension to a discrete language''}, while we aim at the opposite limit of avoiding any discrete language at all. In this initial form, finite temperature field theory is applicable only to some thermal properties of solid matter.

It is interesting to recall a physical theory called the '5-optics' by  Yuri Rumer \cite{Rumer-book1956}. His theory was based on the ideas of Kaluza-Klein and Einstein-Bergman about extra dimensions of spacetime and introduced the closed fifth dimension. This compactified dimension had physical dimensionality of an action with a period equal to the Planck's constant \cite{Rumer-JETP1949}. This is only a curious similarity related to the present work, but the Rumer's works should be useful for researchers in the extra dimensional physics. 

The heat kernel trace is used as a mathematical basis, and its advantage is that it can be computed at once for many relevant field theories, \cite{CPT2,Gusev-NPB2009}. Only after taking the proper time integral (\ref{Fbeta}), one obtains the effective action or free energy, by this step going from geometry to physics. This equation is taken as a definition of the quantum functional. 

In the heat kernel method, finite temperature  QFT looks similar to thermodynamics.  Classical thermodynamics preceded the quantum theory and served a source of ideas and a technical tool for developing the quantum theory by Max Planck \cite{Planck-book1914,Muller-AP2008} and Albert Einstein \cite{Einstein-photon1905,Pais-RMP1979}. The deep relation between statistical physics and quantum theory has been explored extensively starting from the pioneering work of J.E. Moyal \cite{Moyal-MPCRS1949}. The present work looks at the thermodynamics side of this relation from the field theory viewpoint.

%%%%%%%%%%%%%%%%%%%%%%%%%%%%%%%%%%%%%%%%
\section*{Acknowledgment.}
%%%%%%%%%%%%%%%%%%%%%%%%%%%%%%%%%%%%%%%%
The author would like to thank  CERN Theory Division and Perimeter Institute for Theoretical Physics for hospitality and support during the corresponding visits. This work was supported in 2012 by a visiting grant from FAPEMIG (Brasil). The long-term workshop YITP-T-12-03 "Gravity and Cosmology 2012" at Kyoto University is gratefully acknowledged. 

%%%%%%%%%%%%%%%%%%%%%%%%%%%%%%%%%%%

%%%%%%%%%%%%%%%%%%%%%%%%%%%%%%%%%%
\end{document}